\documentclass[12pt]{iopart}
\usepackage{setstack}
\usepackage{verbatim}
\usepackage{bm}
\usepackage{graphicx}
\usepackage{color}
\usepackage{amssymb}

\begin{document}

\newcommand{\ket}[1]{| #1 \rangle}
\newcommand{\bra}[1]{\langle#1 |}

\title{Entanglement evolution of two remote and non-identical Jaynes-Cummings atoms}
\author{Stanley Chan$^{1}$, M D Reid$^{2}$ and Z Ficek$^{1}$}
\address{$^{1}$Department of Physics, School of Physical Sciences, The University
of Queensland, Brisbane, QLD 4072, Australia}
\address{$^{2}$ARC Centre of Excellence for Quantum-Atom Optics, School of
Physical Sciences, The University of Queensland, Brisbane QLD 4072, Australia}

\date{\today}

\begin{abstract}
A detailed treatment of the entanglement dynamics of two distant but
non-identical systems is presented. We study the entanglement evolution
of two remote atoms interacting independently with a cavity field,
as in the double Jaynes-Cummings (J-C) model. The four-qubit pairwise
concurrences are studied, allowing for asymmetric atom-cavity couplings
and off-resonant ineractions. Counter to intuition, imperfect matching
can prove advantageous to entanglement creation and evolution. For
two types of initial entanglement, corresponding to spin correlated
and anti-correlated Bell states $\Phi$ and $\Psi$, a full, periodic
and directed transfer of entanglement into a specific qubit pair is
possible, for resonant interactions, depending on the choice of relative
couplings. Furthermore, entanglement transfer and sudden death (ESD)
can be prevented using off-resonant interactions, although for some
initial states, detunings will trigger an otherwise frozen entanglement,
to allow a full entanglement transfer.
\end{abstract}

\pacs{03.65.Ud, 42.50.Lc, 42.50.Pq}

\maketitle

\section{Introduction}

Entanglement is now regarded to be a resource central to the development
of quantum technologies. lt has been extensively studied theoretically
and creation of entanglement has now been reported in a range of systems,
including trapped ions~\cite{expent}, atomic ensembles~\cite{atomicens}
and photon pairs~\cite{aspecttype}.

A fundamental feature of entanglement is that it is easily degraded
when the entangled system interacts with another~\cite{diosi,dodd,yueberly}.
Work by Yu and Eberly~\cite{yueberly,ebscience} has shown that the
entanglement of two qubits can deteriorate rapidly, to the point of
an abrupt total destruction, when coupled to an environment that results
in irreversible loss. Recent experiments~\cite{alexp,salles} confirm
the sudden death of entanglement (ESD) and there have been further
studies investigating this behaviour~\cite{zt,santos,other esdanna,caval,recent esd terra,sainz,yebx,esd,esd2008}.

Particularly relevant for the purpose of quantum networks are entangled
qubits that can be stored, localised and controlled. Many investigations
have focused on the storing of entanglement in two level atoms, that
are constrained to intertact with a single mode field in a cavity~\cite{qed work exp}.
The use of very high $Q$ cavities and Rydberg atoms~\cite{har-ryd}
minimises the irreversible losses of the type studied by Yu and Eberly.
In this case, the atom interacts only with the single cavity mode
and the Jaynes-Cummings model~\cite{jc,jcexp} is realized. The dynamics
and control of entanglement between two isolated atoms~$A$ and~$B$
in a {}``lattice'', allowed only to interact with two {}``control''
fields $a$ and $b$, repsectively, as in the JC model is thus
of fundamental interest~\cite{yonac,yonac2,yelattice}.

The evolution of the entanglement between two isolated atoms that
are each coupled to a field via Jaynes-Cummings cavity has been studied
recently by Yonac et al~\cite{yonac,yonac2} and Sainz and Bjork~\cite{sainz}.
By examining the pairwise entanglement using the method of concurrences~\cite{wit2},
Yonac et al~\cite{yonac} showed that the atom-atom entanglement
is transferred to the cavity-cavity system, and then back again, in
a periodic fashion. Two types of initial two-qubit entanglement of
the atoms were considered, the Bell state $\Psi$ where the spins
are anti-correlated, and the Bell state $\Phi$ where spins are correlated.
A complete and abrupt loss of atom-atom entanglement (ESD) is found
to occur for~$\Phi$, provided the initial entanglement is not maximum,
though the full entanglement is periodically regained. 

In this paper, we study the effect of non-identical JC systems, so that one
can control the entanglement between $A$ and $B$, via manipulation
of the localised atom-field interaction or detunings. We present complete
entanglement solutions, for the case of asymmetric atom-cavity couplings
and off-resonant interactions (detunings), so that all possible pairwise concurrences
are calculated. The results reveal that a full periodic and directed
transfer of the entanglement from one qubit pair to another occurs
with suitable choice of coupling ratios. We find that the full entanglement
transfer and ESD can be controlled with the use of suitable detunings,
for both initial states $\Psi$ and $\Phi$. Thus the general effect
of detuning is to stabilize entanglement. For some initial states
however, where entanglement is already constant, the effect of detuning
can be to trigger a maximal entanglement.

\section{Hamiltonian and energy states}

We begin by introducing the Hamiltonian of the system and the energy
basis states that we will use to solve for the entanglement creation
and evolution.

\subsection{The Double Jaynes-Cummings Hamiltonian }

\begin{figure}[hbp]
\begin{center}
\includegraphics[clip,width=3in]{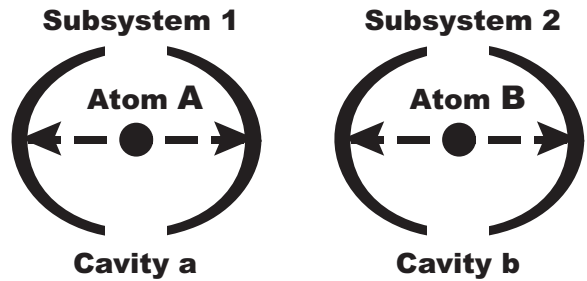} 
\end{center}
\caption{Schematic diagram of the system considered, two distant and non-interacting
subsystems $1$ and $2$ each with a single-mode cavity containing
a two-level atom. 
\label{cap:setup1}}
\end{figure}

We consider a system composed of two separated single mode cavities
each containing a single two-level atom, as shown in Fig.~\ref{cap:setup1}.
In general, the two subsystems may not be identical in that the cavity
frequencies and the coupling strengths of the atoms to the cavity
modes could be different. 

The Hamiltonian for the system in the electric-dipole and rotating-wave
approximations can be written in the Jaynes Cummings (JC) form 
\begin{equation}
\hat{H}=\hat{H}_{Field}+\hat{H}_{Atom}+\hat{H}_{int} ,\label{e1}
\end{equation}
where 
\begin{equation}
\hat{H}_{Field}=\hbar\omega_{1}\left(a^{\dag}a+\frac{1}{2}\right)
+\hbar\omega_{2}\left(b^{\dag}b+\frac{1}{2}\right)\label{e2}
\end{equation}
is the Hamiltonian of the cavity modes with resonant frequencies
$\omega_{1}$ and $\omega_{2}$ respectively,
 \begin{equation}
\hat{H}_{Atom}=\hbar\omega_{0}S_{A}^{z}+\hbar\omega_{0}S_{B}^{z}\label{eq:ha}
\end{equation}
is the Hamiltonian of the atoms of the transition frequency $\omega_{0}$, and 
\begin{equation}
\hat{H}_{int}=\hbar g_{1}\left(a^{\dagger}S_{A}^{-}+aS_{A}^{\dagger}\right)+\hbar g_{2}\left(b^{\dagger}S_{B}^{-}+bS_{B}^{\dagger}\right) \label{eq:hint}
\end{equation}
is the interaction Hamiltonian between the atoms and the cavity modes.
The parameters $g_{1}$ and $g_{2}$ are the strengths of the coupling
between the atoms and the cavity modes, that can be assumed to be
real with no loss of generality. The $S_{i}^{+}$ and $S_{i}^{-}$
are respectively the raising and lowering operators of the $i$th
atom, and $a^{\dag}\ (a)$ and $b^{\dag}\ (b)$ are the creation (annihilation)
operators for the modes of the cavities labelled $a$ and $b$, respectively.

In previous treatments~\cite{yonac,yonac2}, the focus of study has
been on identical and resonant cavities, where $\omega_{1}=\omega_{2}=\omega_{0}$
and $g_{1}=g_{2}$. This makes the two subsystems $1$ and~$2$ non-distinguishable,
which might suggest a greater robustness for entanglement creation.
By contrast, we will show that asymmetric cavities and cavity-atom
detunings can prove advantageous in enabling a control of entanglement.
Our objective then is to include $\omega_{1}\neq\omega_{2}\neq\omega_{0}$
and $g_{1}\neq g_{2}$, which also has the purpose of better modelling
a real experimental situation, where it may be difficult to produce
identical cavities. The unequal coupling constants for example may
arise when atoms are not in equivalent positions inside the cavities. In addition, we find the case of asymmetric couplings leads to a breakdown of a simple entanglement transfer conservation rule derived by Yonac et al~\cite{yonac2}, and proved to hold where $g_1=g_2$.

\subsection{Energy basis states}

We assume an initial entanglement between the distant atomic systems
$A$ and $B$, and study the evolution of the entanglement in the
absence of external losses, where there is a local coupling of each
atom to its cavity mode. The energy basis states of the non-interacting
Hamiltonian are the product states of those for the atoms and the
cavity modes. We denote the energy states of a two-level atom as $|\!\uparrow\rangle$
and $|\!\downarrow\rangle$, for the excited and ground states respectively,
so that $S^{z}|\!\uparrow\rangle=\frac{1}{2}|\!\uparrow\rangle$ and 
$S^{z}|\!\downarrow\rangle=-\frac{1}{2}|\!\downarrow\rangle$.
The energy states of a cavity are denoted by the Fock states $|n\rangle$, with
$n$ being the number of quanta in the cavity mode. The size of the
Hilbert space for the system depends on the number of energy quanta
involved in the evolution of the atoms and the cavity modes. 

We follow the approach of Yonac et al.~\cite{yonac}, and consider
two different types of initial entanglement, corresponding to the
Bell state with spins anti-correlated
 \begin{equation}
|\Psi\rangle=\cos\alpha|\!\uparrow\downarrow00\rangle
+e^{i\beta}\sin\alpha|\!\downarrow\uparrow00\rangle,\label{eq:bell1}
\end{equation}
and the Bell state with spins correlated 
\begin{equation}
|\Phi\rangle=\cos\alpha|\!\uparrow\uparrow00\rangle
+e^{i\beta}\sin\alpha|\!\downarrow\downarrow00\rangle.\label{eq:bell2}
\end{equation}
Here, $|\!\uparrow\downarrow ij\rangle$ represents
the atom $A$ in the excited ({}``up'') state, the atom $B$ in
the ground ({}``down'') state, and $i$ and $j$ photons in the
cavity modes $a$ and $b$, respectively.  We will use the notation $\Psi$ and $|\Psi\rangle$ interchangeably, to represent the quantum Bell state. Following Yonac et al.~\cite{yonac},
we allow for an initial non-maximal entanglement state by incorporating
the parameter $\alpha$.

Firstly, in Section~\ref{sec3}, we consider the initial
state $|\Psi\rangle$ where there is only a single excitation present,
so that the space of the system is spanned by four state vectors,
defined as follows:
 \begin{eqnarray}
|\Psi_{1}\rangle  =  |\!\uparrow\downarrow00\rangle ,\quad 
|\Psi_{2}\rangle  =  |\!\downarrow\uparrow00\rangle ,\quad
|\Psi_{3}\rangle  = |\!\downarrow\downarrow10\rangle ,\quad 
|\Psi_{4}\rangle  =  |\!\downarrow\downarrow01\rangle .\label{e5}
\end{eqnarray}
In addition, we will study the {}``delocalised'' entangled states
of type $\sum_{i=1}^{4}|\Psi_{i}\rangle/2$, where the component states
have equal weightings, to show that such states can give a stable,
albeit reduced, entanglement.

Secondly, in Section~\ref{sec4}, we will consider the case where there are
two excitations present, one in each system, so the space of the system
is spanned by the vectors 
\begin{eqnarray}
|\Phi_{1}\rangle  =  |\!\uparrow\uparrow00\rangle ,\quad 
|\Phi_{2}\rangle  =  |\!\uparrow\downarrow01\rangle ,\quad
|\Phi_{3}\rangle  =  |\!\downarrow\uparrow10\rangle ,\quad 
|\Phi_{4}\rangle  =  |\!\downarrow\downarrow11\rangle.\label{eq:states2}
\end{eqnarray}
In this case we also include the ground state of the system
\begin{equation}
|0\rangle=|\Phi_{0}\rangle=|\downarrow\downarrow00\rangle\label{eq:ground}
\end{equation}
for which there is no excitation present, so that we can specifically
study the case where the initial entanglement is in the form of the Bell state~(\ref{eq:bell2}).

We will quantify entanglement by using the measure of concurrence,
introduced by Wootters~\cite{wit2}. The concurrence is defined as
\begin{equation}
C={\rm max}\left(0,\sqrt{\lambda_{1}}-\sqrt{\lambda_{2}}-\sqrt{\lambda_{3}}
-\sqrt{\lambda_{4}}\right),\label{eq:concurrence}
\end{equation}
where $\lambda_{i}$ are the eigenvalues of the density matrix $\rho'=\rho(\sigma_{y}\otimes\sigma_{y})\rho^{\ast}(\sigma_{y}\otimes\sigma_{y})$,
and $\sigma_{y}$ is the Pauli matrix in $y$ direction. The maximum
possible entanglement is given by $C=1$, while $C=0$ implies separability.
The state is entangled (inseparable) for $C>0$. One can calculate that the concurrence for the states $|\Psi\rangle$  and $|\Phi\rangle$  is 
\begin{equation}
 C_{12}=2\sin\alpha\cos\alpha,\label{eq:concurrencebell}
 \end{equation}
to indicate maximal entanglement for $\alpha=\pi/4$.

The transfer of entanglement between the two qubit subsystems is governed by rules derived recently  by Yonac et al.~\cite{yonac2}, and Chan et al.~\cite{smznew}. Sainz and Bjork~\cite{sainz} have also presented an entanglement invariant, for the case of the double JC model. Chan et al.~\cite{smznew} have proved the following result for entanglement initially in the form of the Bell state $|\Psi\rangle$:
 \begin{equation}
C_{AB}^{2}+C_{Ab}^2+C_{aB}^2+C_{ab}^2=C_{12}^{2}.\label{eq:concurrenceequal}
\end{equation}
The total non-local pairwise entanglement, as measured by the concurrence squared, is conserved, where $C_{12}$  is defined as in Eq.~(\ref{eq:concurrencebell}) to be the bipartite entanglement between the systems we label  $1$ and $2$. In fact, from the work of Chan et al., it is seen that this rule applies to any initial state for which there is a single-excitation (refer to Section~\ref{sec3}). For the case of symmetric JC couplings, where $g_{1}=g_{2}$, Yonac et al~\cite{yonac2} have proved that in fact the following simpler result also holds, if the atoms are initially in the Bell state $|\Psi\rangle$: 
\begin{equation}
C_{AB}+C_{ab}=C_{12}.\label{eq:yonac}
\end{equation}
For the case of entanglement initially in the form of the Bell state $|\Phi\rangle$, the pairwise entanglement is not conserved~\cite{lopez}, and following inequality is always valid~\cite{smznew} for non-interacting systems $1$  and $2$:
 \begin{equation}
0\leq C_{AB}^{2}+C_{Ab}^2+C_{aB}^2+C_{ab}^2 \leq C_{12}^{2}.\label{eq:concurrenceunequal}
\end{equation}

\section{Evolution of single-excitation states}\label{sec3}

\subsection{Dynamical solutions}

We consider first the dynamics of the system in a pure state where
only one excitation is present, so that the state of the system is
always of the form 
\begin{equation}
|\Psi_{T}(t)\rangle=\cos\alpha|10\rangle+e^{i\beta}\sin\alpha|01\rangle.\label{eq:entcons}
\end{equation}
The entanglement properties of such a state for the case of perfect
coupling have been studied in detail by Yonac et al.~\cite{yonac,yonac2}.
We allow for the possibility of imperfect couplings of the atoms to
the cavity modes, where detunings can be non-zero and coupling constants
different. For the purpose of solution, we introduce the wave function
of the system 
\begin{eqnarray}
|\Psi(t)\rangle =  d_{1}(t)|\uparrow\downarrow00\rangle+d_{2}(t)|\downarrow\uparrow00\rangle
 +d_{3}(t)|\downarrow\downarrow10\rangle+d_{4}(t)|\downarrow\downarrow01\rangle,\label{e7}
\end{eqnarray}
which is a linear combination of the available product states~(\ref{e5}) of the atoms and the cavity modes at time~$t$. The coefficient $d_{i}(t)$ determines the probability amplitude of the $i$th state
at time~$t$.

We find the coefficients $d_{i}(t)$ by solving the Schrdinger equation
\begin{equation}
i\hbar\frac{d}{dt}|\phi(t)\rangle=\hat{H}|\phi(t)\rangle.\label{e8}
\end{equation}
 It is easily verified that the coefficients satisfy the differential
equations 
\begin{eqnarray}
\dot{d_{1}} & = & -ig_{1}d_{3},\quad\dot{d_{2}}=-ig_{2}d_{4},\nonumber \\
\dot{d_{3}} & = & 2i\Delta_{1}d_{3}-ig_{1}d_{1},\quad\dot{d_{4}}=2i\Delta_{2}d_{4}-ig_{2}d_{2},\label{e9}\end{eqnarray}
 where $2\Delta_{j}=(\omega_{0}-\omega_{j})$ is the detuning of the
$j$th cavity frequency from the atomic transition frequency.

Equations (\ref{e9}) form decoupled pairs of differential equations
that can be easily solved by using e.g. the Laplace transform technique.
A simple solution of the equations, valid for an arbitrary initial
state is given by
\begin{eqnarray}
d_{1}(t) & = & \frac{e^{i\Delta_{1}t}}{\Omega_{1}}\left\{ \Omega_{1}d_{1}(0)\cos(\Omega_{1}t)
-i\left[\Delta_{1}d_{1}(0)+g_{1}d_{3}(0)\right]\sin(\Omega_{1}t)\right\} ,\nonumber \\
d_{3}(t) & = & \frac{e^{i\Delta_{1}t}}{\Omega_{1}}\left\{ \Omega_{1}d_{3}(0)\cos(\Omega_{1}t)
+i\left[\Delta_{1}d_{3}(0)-g_{1}d_{1}(0)\right]\sin(\Omega_{1}t)\right\} ,\label{e11}
\end{eqnarray}
where $\Omega_{1}=\sqrt{g_{1}^{2}+\Delta_{1}^{2}}$ is a detuned
Rabi frequency, $d_{1}(0)$ and $d_{3}(0)$ are the initial values
of the probability amplitudes. Corresponding expressions for the coefficients~$d_{2}(t)$ 
and~$d_{4}(t)$, which determine the evolution of second
cavity system, are obtained from~(\ref{e11}) by exchanging $1\rightarrow2$
and $3\rightarrow4$, and replacing $g_{1}$ by $g_{2}$, and $\Delta_{1}$
by~$\Delta_{2}$.

The probability amplitudes oscillate sinusoidally with the Rabi frequency
$\Omega_{1}$, and their dynamics is strongly affected by the modulation
term that depends on the detuning~$\Delta_{1}$ and the coupling strength
$g_{j}$ between the atom and the corresponding cavity mode. These
modulation terms vanish periodically at time $t_{n}=n\pi/\Omega_{1}$.
It is worth noting that the detuning enters the solutions in an antisymmetric
way, whereas the coupling strength enters the solutions in a symmetric
way. This difference will be evident in the features of the time evolution
of entanglement in the system. It is also worth pointing out that
each of the amplitudes is influenced by the initial value of the other.
This fact will also have a crucial effect on the evolution of an entanglement
in the system.

Solutions (\ref{e11}) differ from previous results of Ref.~\cite{yonac2}
in that non-zero detunings and unequal coupling strengths are included. Non-vanishing
detunings and differences between the coupling strengths lead to new
aspects of the entanglement creation and evolution that differ qualitatively
and quantitatively from those observed under exact resonance and equal
couplings.

\subsection{Single-excitation entanglement concurrences}

Our interest is centered principally on the evolution of entanglement
between the different parts of the two two-qubit subsystems $1$ and
$2$. By denoting the two atoms as $A$ and $B$, and the corresponding
cavity modes as $a$ and $b$, we may distinguish six pairs of sub-systems
$AB,ab,Aa,Ab,Ba$ and $Bb$, as investigated by Yonac et al~\cite{yonac,yonac2}. 

To calculate the entanglement between any two qubit pair $IJ$, for
example the atoms $A$ and $B$, we take the trace over the other
subsytems, to evaluate the reduced density matrix, from which the
eigenvalues can be calculated. In general, the reduced states are
two qubit mixtures, and have a density matrix of the general {}``$X$-form''~\cite{yebx,esd} 
\begin{equation}
\rho_{IJ}=\left(\begin{array}{cccc}
a & 0 & 0 & z\\
0 & b & 0 & 0\\
0 & 0 & c & 0\\
z^{*} & 0 & 0 & d\end{array}\right),\label{eq:xmtrix}
\end{equation}
which has the concurrence
\begin{equation}
C_{IJ}=2\max\bigl\{0,|z|-\sqrt{bc}\bigr\}.\label{eq:xconcurr}
\end{equation}

We solve for all the two qubit concurrences, $C_{AB},C_{ab},C_{Aa},C_{Bb},C_{Ab},C_{Ba}$,
and find that in terms of the probability amplitudes, their time evolutions
are given by the simple expressions
\begin{eqnarray}
C_{AB}(t) & = & 2|d_{1}(t)||d_{2}(t)|,\, C_{ab}(t)=2|d_{3}(t)||d_{4}(t)|\nonumber \\
C_{Aa}(t) & = & 2|d_{1}(t)||d_{3}(t)|,\, C_{aB}(t)=2|d_{2}(t)||d_{3}(t)|\nonumber \\
C_{Ab}(t) & = & 2|d_{1}(t)||d_{4}(t)|,\, C_{Bb}(t)=2|d_{2}(t)||d_{4}(t)|.\label{eq:conds} 
\end{eqnarray}
There are several remarks that can be made about the general results
for the concurrence measures, before we proceed to the detailed analysis
of the entanglement evolution. In the first place, we observe that
an entanglement between the subsystems $1$ and $2$ cannot be created
in time if initially the excitation was entirely in the $1$ or $2$
subsystem, so that $C_{AB}(t),C_{ab}(t),C_{Ab}(t)$, $C_{Ba}(t)$
are all initially zero. Since the subsystems $1$ and $2$ are not
directly coupled to each other, no transfer of the excitation, and
hence entanglement, is possible during the evolution. This means that
an initial entanglement between the subsystems must be imposed to
observe any entanglement evolution.

Secondly, we note that an entanglement between $A$ and $a$ or between
$B$ and $b$ can be created in time if the excitation is either in
the $1$ or $2$ subsystem. This is easy to understand. Since the
atoms interact with the cavity modes to which they are coupled, they
can share the excitation with the corresponding mode. 

Finally, if initially $C_{Aa}=1$ or $C_{Bb}=1$, to indicate maximum
possible entanglement between each atom and its local cavity mode,
one can easily show that the entanglements remain constant (frozen)
in time unless the detunings $\Delta_{1}$ or $\Delta_{2}$ are different
from zero. In this case, the entanglement measures evolve in time,
but cannot be zero.

\subsection{Initial states that result in a {}``frozen'' pairwise entanglement evoultion}

\begin{figure}[hbp]
\begin{center}
\includegraphics[clip,width=4in]{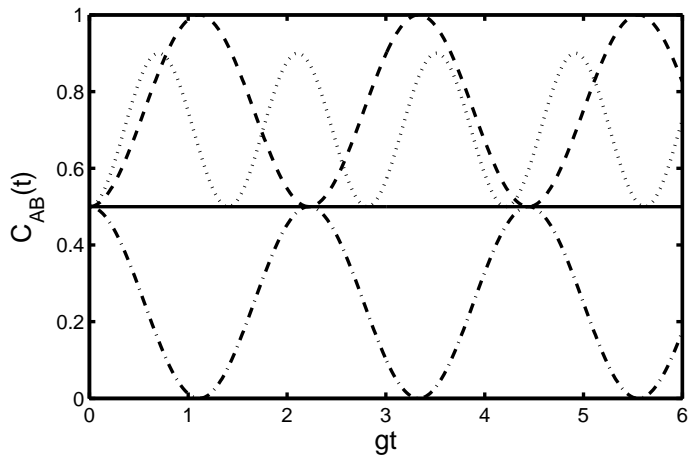} 
\end{center}
\caption{Time evolution of the concurrence measure $C_{AB}$ showing stability
of entanglement between atoms for zero detunings $\Delta_{1}=\Delta_{2}=0$,
and arbitrary $g_{1}$, $g_{2}$, for the intial state (\ref{e15})
(solid line). The dashed lines reveal the effects of detuning, where
$g_{1}=g_{2}=g$: $\Delta_{1}=\Delta_{2}=g$ (dashed line), $\Delta_{1}=\Delta_{2}=-g$
(dashed-dotted line), $\Delta_{1}=\Delta_{2}=2g$ (dotted line). 
\label{cap:syment2}}
\end{figure}

There is the question of which initial states have stable, or {}``frozen'',
entanglement properties under the evolution of the J-C Hamiltonian.
We consider the superposition with a uniform population distribution
over the available energy states (\ref{e5}). We write
\begin{eqnarray}
|\Psi_{0}\rangle=\frac{1}{2}\left(|\Psi_{1}\rangle
+e^{i\theta}|\Psi_{3}\rangle\pm(|\Psi_{2}\rangle-e^{i\phi}|\Psi_{4}\rangle)\right),\label{e15}
\end{eqnarray}
where $\theta$ and $\phi$ are arbitrary phase factors. For these
states, in the case of zero detunings, we see from~(\ref{eq:conds})
that all of the two-qubit entanglement between pairs remains constant
with time. In this case, the entanglement is maximally shared (we call this {}``delocalized")
between the six subsystem pairs, so $C_{AB}(0)=C_{Ab}(0)=....=C_{Bb}(0)=1/2$
(see Fig.~\ref{cap:syment2}), in accordance with the rule  ~(\ref{eq:concurrenceequal}). Thus, we see that where detunings are zero 
and for symmetric coupling constants $g_{1}=g_{2}$, the superposition state (\ref{e15})
is a stationary state of the system, and hence must maintain fixed
entanglement features. Interestingly though, this stability of all
two-qubit entanglement concurrences is also predicted for asymmetric
couplings, where $g_{1}\neq g_{2}$.

The effect of detunings $\Delta$ on the time evolution
of the atom-atom concurrence $C_{AB}(t)$ is quite different. For
the initial state (\ref{e15}), the evolution of concurrence $C_{AB}$
is plotted in Fig.~\ref{cap:syment2} for equal coupling constants
$g_{1}=g_{2}=g$. The atomic concurrence varies in time only for
nonzero detuning and the time evolution of the concurrence is not
symmetric with respect to the sign of the detunings. The dependence
on time of the atom-atom concurrence $C_{AB}(t)$ changes qualitatively
as well as quantitatively when the detunings $\Delta$ are varied
from positive to negative values. A large and even \emph{maximal}
entanglement between the atoms can be created when the detunings are
\emph{positive}, whereas the initial atomic entanglement is reduced
and can even be suppressed when the detunings are \emph{negative}.
We note that from the equality~(\ref{eq:concurrenceequal}),
a maximum entanglement ($C_{AB}=1$) between atoms will imply zero
entanglement between the two cavities. When the atom-atom entanglement
is maximum, at $C_{AB}=1$, the cavity-cavity entanglement is zero,
at $C_{ab}=0$, and vice versa, so that this result also implies a
channeling into the cavity-cavity system. In addition, an important
result is that the maximum atomic entanglement $C_{AB}=1$ is created 
when the ratios $\Delta_{1}/g_{1}=\Delta_{2}/g_{2}=1$.
Otherwise, the atomic entanglement is reduced. 

The effect of detuning implies that one can engineer the direction
of evolution of entanglement by controlling the detuning. By this
we mean that an initial {}``non-localized'' fully symmetric entanglement
and population, as given by~(\ref{e15}) can be transferred to
a desired {}``localized'' atom-atom entanglement and population
by a suitable choice of the detunings: positive detuning can channel
entanglement entirely into the atoms, or cavities, at appropiate times.
{}``Localized'' in this context means that the entanglement exists
solely between the two atoms, or between the two fields, as in~(\ref{eq:bell1}).

\subsection{Entanglement evolution for the Bell state $\Psi$}

Next, we consider the evolution of the concurrences for the following
Bell initial state 
\begin{eqnarray}
|\Psi\rangle = \frac{1}{\sqrt{2}}\left(|\Psi_{1}\rangle\pm|\Psi_{2}\rangle\right)
 =  \frac{1}{\sqrt{2}}\left(|\uparrow\downarrow00\rangle\pm|\downarrow\uparrow00\rangle\right) ,
 \label{e15a}
 \end{eqnarray}
where the entanglement is restricted to just the two atoms.
 
Here the population is solely localized in the atoms, and the entanglement
of the pair is maximum, corresponding to $C_{AB}=1$. The time evolution
of the concurrence measures is given by 
\begin{eqnarray}
C_{AB}(t) & = & \bigl|\cos\left(\Omega_{1}gt\right)\bigl|\bigl|\cos\left(\Omega_{2}gt\right)\bigl|,\quad 
C_{ab}(t)  = \bigl|\sin\left(\Omega_{1}gt\right)\bigl|\bigl|\sin\left(\Omega_{2}gt\right)\bigl|,\nonumber \\
C_{Aa}(t) & = & \bigl|\cos\left(\Omega_{1}gt\right)\bigl|\bigl|\sin\left(\Omega_{1}gt\right)\bigl|,\quad
C_{Ab}(t)  =  \bigl|\cos\left(\Omega_{1}gt\right)\bigl|\bigl|\sin\left(\Omega_{2}gt\right)\bigl|,\nonumber \\
C_{Ba}(t) & = & \bigl|\cos\left(\Omega_{2}gt\right)\bigl|\bigl|\sin\left(\Omega_{1}gt\right)\bigl|,\quad
C_{Bb}(t)  =  \bigl|\sin\left(\Omega_{2}gt\right)\bigl|\bigl|\cos\left(\Omega_{2}gt\right)\!\bigl|,\label{e16}
\end{eqnarray}
for the case of exact resonances $\Delta_{1}=\Delta_{2}=0$ and unequal
coupling strengths, $g_{1}\neq g_{2}$. Here $\Omega_{1}=(1+u/g)$
and $\Omega_{2}=(1-u/g)$ are dimensionless Rabi frequencies,
with $g=(g_{1}+g_{2})/2$ and $u=(g_{1}-g_{2})/2$. 

For equal coupling strengths $g_{1}=g_{2}\equiv g$ and non-zero detunings,
the bipartite concurrence measures are \begin{eqnarray}
C_{AB}(t) & = & \cos^{2}\left(\Omega gt\right)+\left(\frac{\delta}{\Omega}\right)^{2}\sin^{2}\left(\Omega gt\right),\nonumber \\
C_{ab}(t) & = & \frac{1}{\Omega^{2}}\sin^{2}\left(\Omega t\right),\nonumber \\
C_{Aa}(t) & = & C_{Ab}(t)=C_{Ba}(t)=C_{Bb}(t)=\frac{1}{\Omega}\sin\left(\Omega gt\right)\nonumber \\
 &  & \times\left[\cos^{2}\left(\Omega gt\right)+\left(\frac{\delta}{\Omega}\right)^{2}\sin^{2}\left(\Omega gt\right)\right]^{1/2},\label{e17}\end{eqnarray}
 where we have simplified the detunings to $\Delta_{1}=\Delta_{2}\equiv\Delta$
and have introduced a dimensionless Rabi frequency, the same for both
subsy\textcolor{black}{stems, $\Omega=(1+\delta^{2})^{1/2}$ with $\delta=\Delta/g$.}
\begin{figure}[hbp]
\begin{center}
\includegraphics[clip,width=4in]{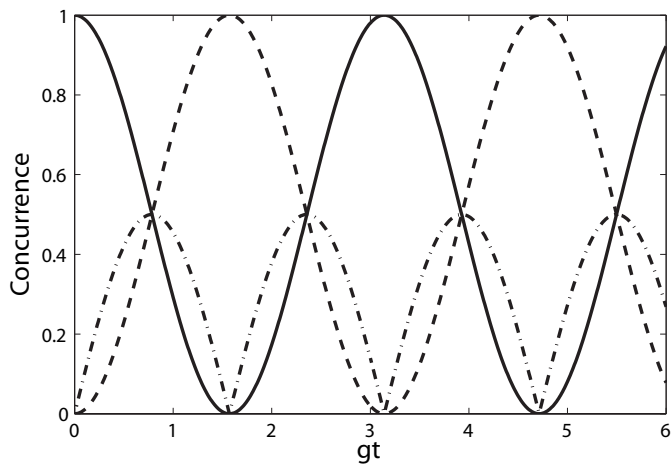} 
\end{center}
\caption{Oscillation of the concurrence measures for equal coupling strengths,
$g_{1}=g_{2}=g$ and exact resonance $\Delta_{1}=\Delta_{2}=0$, for
the initial Bell state $\Psi$, given by~(\ref{e15a}). $C_{AB}(t)$ (solid
line), $C_{ab}(t)$ (dashed line), and $C_{Aa}(t),C_{aB}(t),C_{Ab}(t),C_{bB}(t)$
(dashed-dotted line). Full transfer of entanglement from atoms to
cavity modes takes place\textcolor{black}{\ at $t=n\pi/2\Omega$,
where $n=1,3,5....$. }, \textcolor{black}{in accordance with both
conservation rules~(\ref{eq:concurrenceequal}) and~(\ref{eq:yonac}).
Only~(\ref{eq:concurrenceequal}) holds for asymmetric $g$'s
(see Fig. (\ref{cap:conservlawsdiffg4}))}.
\label{cap:concosc3}}
\end{figure}

These expressions are plotted in Fig. \ref{cap:concosc3}
to reveal the periodic transfer of entanglement from atoms to cavity
modes, and vice versa. The oscillatory behaviour of the atom-atom
concurrence was reported by Yonac et al.~\cite{yonac2}. We note the
transfer process of entanglement from the atoms to the cavity modes
does not involve just the pairs $C_{AB}$ and $C_{ab}$. By inspection
of the time evolution of the concurrence measures in Fig. \ref{cap:concosc3},
we find that the initial maximal entanglement between the atoms is
not only totally transferred to the cavity modes, but at the same
time an additional pairwise entanglement is created during the evolution.
This can also be seen by summing the pair concurrence measures to
find that at times $t_{n}=n\pi/4\Omega$, where $n=1,3,5...$, the
total pair entanglement is larger than one. As time progresses, a
part of the entanglement is transferred into the other pairs of the
subsystems. Then, after a further interval, the entanglement is completely
transferred into the cavity modes. The additional entanglement vanishes
at times $t_{n}=n\pi/2\Omega$, $n=1,2,3,...$, when the transfer
process is completed.

\begin{figure}[hbp]
\begin{center}
\includegraphics[clip,width=4in]{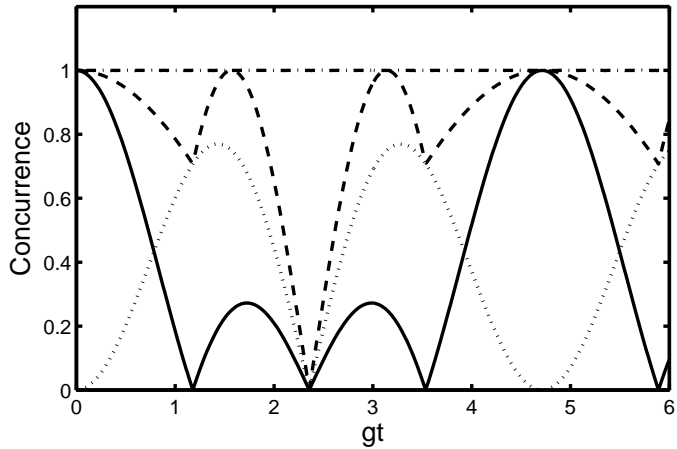} 
\end{center}
\caption{Time evolution for the single-excitation state, showing the conservation
of the sum of the individual pairwise $1-2$ concurrences (SSPC): $C_{AB}$
(solid line); $C_{ab}$ (dottted line); $C_{12}$ (dashed-dotted line);
$C_{AB}+C_{ab}$ (dashed line). The total $C_{AB}^{2}+C_{ab}^{2}+C_{Ab}^{2}+C_{aB}^{2}$
is equal to $C_{12}^{2}=1$ as indicated by~(\ref{eq:concurrenceequal}).
\textcolor{black}{Here $\Delta=0$, $g_{1}=2g_{2}$, $g=(g_{1}+g_{2})/2$
} and \textcolor{black}{$\alpha=\pi/4$. We note that in this case,
$C_{AB}+C_{ab}$ is not constant, and the conservation rule~(\ref{eq:yonac})
breaks down. }
\label{cap:conservlawsdiffg4}}
\end{figure}

We note however that the nonlocal entanglement, as defined by ~(\ref{eq:concurrenceequal}) is conserved throughout the transfer process. For the case of $g_1=g_2$ the rule  ~(\ref{eq:yonac}) also applies, as is evident in Fig. \ref{cap:concosc3}. Where there is asymmetry in the interaction, so we have non-identical cavities, this rule breaks down, though the general conservation rule  ~(\ref{eq:concurrenceequal}) still applies. This is evident in Fig. \ref{cap:conservlawsdiffg4} and has been noted by Chan et al.~\cite{smznew}. 

\begin{figure}[hbp]
\begin{center}
\includegraphics[clip,width=4in]{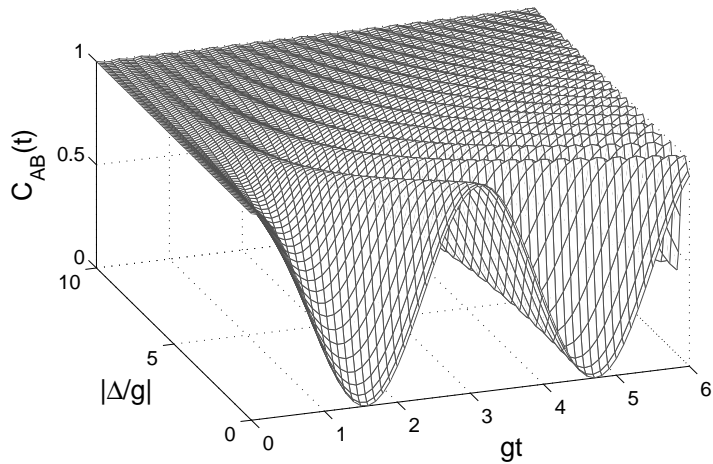} 
\end{center}
\caption{Effect of detuning on the entanglement evolution of the Bell state
$\Psi$, given by~(\ref{e15a}). The concurrence measure $C_{AB}$ is plotted
as a function of time and detuning $\Delta$. Here
$g_{1}=g_{2}=g$. Detuning stabilises the entanglement between atoms
and decreases the period of entanglement revival. 
\label{cap:detuningcon5}}
\end{figure}

\begin{figure}[hbp]
\begin{center}
\includegraphics[clip,width=4in]{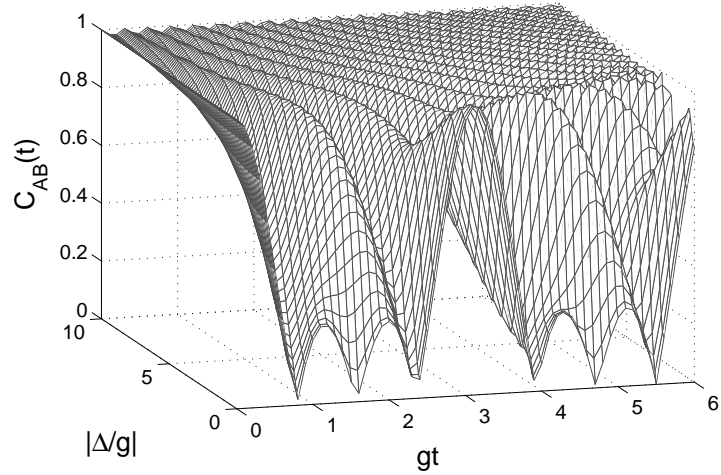} 
\end{center}
\caption{Effect of detuning on the evolution of the atom-atom concurrence $C_{AB}$,
for asymmetric $g$'s, and where the initial state is the Bell state
$\Psi$, given by~(\ref{e15a}). Here $g_{1}=2g_{2}$,
and $g=(g_{1}+g_{2})/2$. Effect is similar to the case of symmetric
$g$'s, except for the modulation of the period of oscillation. 
\label{cap:detasymg}}
\end{figure}

\begin{figure}[hbp]
\begin{center}
\includegraphics[clip,width=4in]{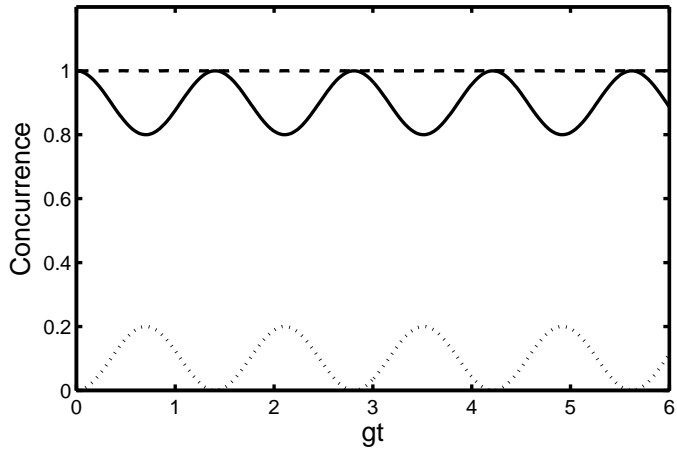} 
\end{center}
\caption{Time evolution of the concurrences $C_{AB}$ (solid line), $C_{ab}$
(dotted line\textcolor{black}{)} for the Bell state $\Psi$, showing
the effect of detuning. The transfer of entanglement from atoms to
cavities is not complete, but a full periodic transfer back to atoms
occurs.\textcolor{black}{  Here, $\Delta=2g$ and $g_{1}=g_{2}=g$.
We note that in this case, the simple conservation rule~(\ref{eq:yonac})
of Yonac et al ~\cite{yonac2} holds.}
\label{cap:detuningcavitatomconc}}
\end{figure}

Another interesting prediction of~(\ref{e16}) is that complete entanglement transfer from the atoms to the cavity fields requires exact resonances but not necessarily equal coupling
constants. To examine the effect of detuning first, in Figures \ref{cap:detuningcon5}
and \ref{cap:detasymg}, we note that where detunings $\Delta_{1}$
and $\Delta_{2}$ are nonzero, the localized initial entanglement
is not completely transferred to another pair of qubits (see Fig.~\ref{cap:detuningcavitatomconc}). Detuning increases the oscillation
frequency and decreases the minimum entanglement between the atoms.
This is in complete contrast to the case of the initially completely
symmetric entanglement, where entanglement transfer
is possible only for a nonzero detuning. It is interesting, however,
that the atomic entanglement returns to its initial maximum value
periodically with the detuned Rabi frequency. This can be understood
in terms of the localization of the initial energy. When $\Delta\neq0$,
only a part of the initial energy, proportional to $\omega_{0}$ is
transferred to the cavity modes leaving the excess energy unlocalized.
We illustrate this situation in Fig. \ref{cap:detuningcon5}, where
we plot the concurrence measure $C_{AB}(t)$ as a function of time
and the detuning $\Delta\equiv\Delta_{1}=\Delta_{2}$, and note that
a nonzero detuning can thus be used to stabilise, or {}``freeze'',
the entanglement between the atoms. 

A second advantage to be given by imperfectly matched
and asymmetric systems lies in the use of asymmetric couplings. We
examine the effect of $g_{1}\neq g_{2}$, which has in part been studied
by Sainz and Bjork~\cite{sainz} and Cavalcanti et al~\cite{caval}.
These authors reveal that a periodic revival of full entanglement
between atoms is possible, as well as a full transfer of entanglement
to cavity systems. By studying all the pairwise concurrences, we report
the new result, that for cavities with unequal rather then equal coupling
strengths, one can engineer the transfer process so that the initial
entanglement is fully transferred into a desired qubit pair. It is
of practical importance to know if an initial localized entanglement
can be transferred on demand with the perfect fidelity to a particular
pair of qubits. Consider the time evolution of the initial atomic
Bell state $\Psi$, as given by~(\ref{e15a}). In the case of
unequal coupling strengths, the entanglement evolution is described
by~(\ref{e16}) from which one can easily find that if the ratio
$n_{g}=g_{1}/g_{2}$ is not an integer number or a fraction of an
integer number, no complete transfer of the state~(\ref{e15a})
is possible to any of the qubit pairs. The complete transfer is possible
\textit{only}  if $n_{g}$ is an integer number or a fraction of an integer number. However,
the destination to where the initial entanglement can be completely
transfered depends on whether $n_{g}$ is an even or an odd integer number.

\begin{figure}[hbp]
\begin{center}
\includegraphics[clip,width=4in]{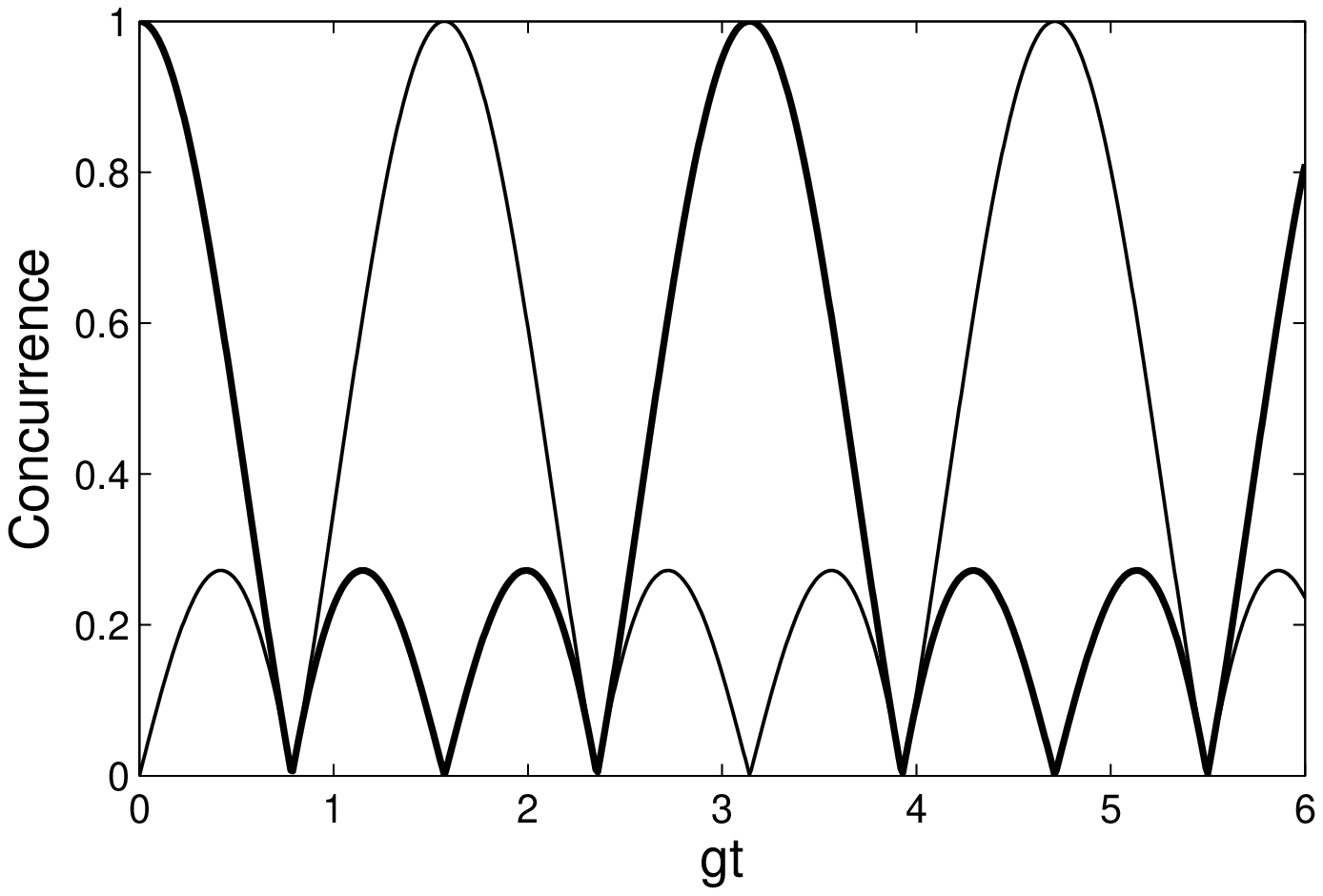} 
\end{center}
\caption{Effect of asymmetric coupling constants on the evolution of the entanglement
of the initial Bell state $\Psi$, given by~(\ref{e15a}).
Concurrence measures are evaluated for resonance $\Delta_{1}=\Delta_{2}=0$
and $g_{1}=2g_{2}$: $C_{AB}(t)$ (solid line), $C_{Ab}(t)$ (sli\textcolor{black}{m
line). In this case, maximum entanglement is transferred periodically
from atoms $AB$ to the atom-field system $Ab$}. 
\label{cap:asymcoupling2fig6}}
\end{figure}

If $n_{g}$ is an even number, the initial maximal
entanglement between the atoms, $C_{AB}(0)=1$, can be completely
transferred only to the atom-field qubit pair $C_{\alpha b}$. On
the other hand, if $n_{g}$ is a fraction of an even number, the initial
entanglement between the atoms can be completely transferred only
to the atomic-field qubit pair $C_{Ba}$. We illustrate this situation
in Fig.~\ref{cap:asymcoupling2fig6}, where we plot the concurrence
of the different qubit pairs as a function of time for exact resonances
but unequal coupling strengths with $n_{g}=2$. It is evident from
the figure that at particular discrete times, the initial entanglement
between the atoms is completely transferred to the qubit pair $C_{Ba}$.
The transfer of the entanglement to other pairs of qubits is incomplete
at all times, the concurrences oscillate between zero and certain
values below unity. The reason for this feature of the entanglement
transfer can be understood intuitively by noting that for $n_{g}=2$
the Rabi frequency $\Omega_{1}$ of the population oscillation in
the subsystem $1$ is twice that of the Rabi frequency $\Omega_{2}$
for the population oscillation in the subsystem $2$. This means that
over a complete Rabi cycle $\Omega_{1}t=\pi$, the initial population
in the subsystem $1$ returns to the atom, but at the same time the
population makes a half Rabi cycle in the subsystem $2$, i.e. the
excitation in system $2$ will be in the cavity mode. Thus, $C_{Ab}=1$
at that time, with the concurrence in the other qubit pairs equal
to zero.
\begin{figure}[h]
\begin{center}
\includegraphics[clip,width=4in]{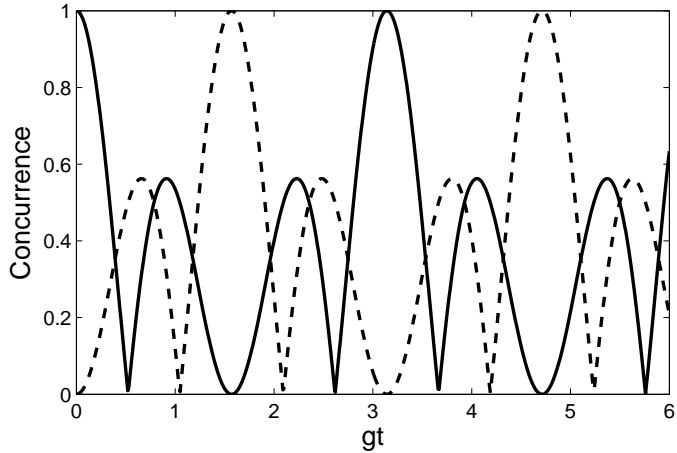} 
\end{center}
\caption{Effect of asymmetric coupling constants on the evolution of the entanglement
of the initial Bell state $\Psi$, given by~(\ref{e15a}).
The concurrence measures are evaluated for $\Delta_{1}=\Delta_{2}=0$
and $g_{1}=3g_{2}$: $C_{AB}(t)$ (solid line), $C_{ab}(t)$ (dashed
line). \textcolor{black}{In this case, maximum entanglement is transferred
to the field-field system $ab$. }
\label{fig10}}
\end{figure}

We now consider the situation where the ratio $n_{g}=g_{1}/g_{2}$
is an odd integer number. Like the previous case of even $n_{g}$,
the initial entanglement can be completely transferred to only one
of the qubit pairs, in this case, to the cavity modes. This is illustrated
in Fig. \ref{fig10}, where we plot the time evolution
of the concurrence of the qubit pairs for the exact resonance but
unequal coupling strengths with $n_{g}=3$. During the evolution,
the initial entanglement $C_{AB}(0)=1$ is completely transferred
to the cavity modes, $C_{ab}(t)=1$ at $\Omega_{1}t=2\pi$. Again,
the transfer of the entanglement to only one of the entanglement pairs
can be understood in terms of the Rabi cycles of the population in
each of the subsystems. Since $\Omega_{1}$ is triple $\Omega_{2}$,
over the one and half of a Rabi- cycle in the subsystem $1$, the
population is completely transferred to the cavity mode $a$, and
at the same time, the population in the subsystem $2$ is also completely
transferred to the cavity mode $b$. Consequently, at this time, the
cavity modes become maximally entangled with $C_{ab}(t)=1$.

In concluding this section, we would like to emphasize that we have
found a significant difference between the evolutions of initial {}``fully
symmetric non-localized'' and {}``targeted localized'' entanglements.
That is, we have found that the {}``non-localized'' entanglement
of states~(\ref{e15}) evolves only if the atoms are imperfectly
coupled to the cavity modes, so that detunings $\Delta$ are nonzero.
On the other hand, where entanglement is initially targeted solely
to the atoms, as in~(\ref{e15a}), large detunings provide a means
of stabilising the entanglement so it cannot be completely transferred
to the field system. Moreover, for zero detunings in this case, one
can achieve complete transfer of an initial entanglement to a desired
pair of qubits, this transfer process being steered by changing the
ratio between the coupling strengths $g$ of the atoms to the cavity
modes.

\section{Evolution of two-excitation entanglement}\label{sec4}

We now turn to the problem of entanglement dynamics when two excitations
are can be present. This allows us to consider the evolution for the
case of the Bell state 
\begin{equation}
|\Phi\rangle=\cos\alpha|\uparrow\uparrow00\rangle+e^{i\theta}\sin\alpha|\downarrow\downarrow00\rangle,\label{eq:bell22}
\end{equation}
which has been studied previously~\cite{yonac}, and found to undergo
an abrupt depletion of entanglement for $\alpha\neq\pi/4$, called
{}``sudden death of entanglement'' (ESD). 

Using the Schrdinger equation, we set up equations of motion for
the probability amplitudes and then solve them to study the dynamics
of the entanglement. It is easy to show that the equations of motion
for the probability amplitudes of the states (\ref{eq:states2}) group
into three independent sets of coupled equations. We define amplitudes
for the wavefunction as
\begin{eqnarray}
|\Psi\rangle & = & d_{1}(t)|\uparrow\uparrow00\rangle
+d_{2}(t)|\uparrow\downarrow01\rangle
+d_{3}(t)|\downarrow\uparrow10\rangle+d_{4}(t)|\downarrow\downarrow11\rangle\nonumber \\
 &  & +d_{0}(t)|\downarrow\downarrow00\rangle\label{et1}
 \end{eqnarray}
in terms of the energy basis states defined by~(\ref{eq:states2}). The evolution of this set involves four equations 
\begin{eqnarray}
\dot{\tilde{d}}_{1} & = & -i(\Delta_{1}+\Delta_{2})\tilde{d}_{1}-i(g_{2}\tilde{d}_{2}+g_{1}\tilde{d}_{3}),\nonumber \\
\dot{\tilde{d}}_{2} & = & -i(\Delta_{1}-\Delta_{2})\tilde{d}_{2}-i(g_{2}\tilde{d}_{1}+g_{1}\tilde{d}_{4}),\nonumber \\
\dot{\tilde{d}}_{3} & = & i(\Delta_{1}-\Delta_{2})\tilde{d}_{3}-i(g_{1}\tilde{d}_{1}+g_{2}\tilde{d}_{4}),\nonumber \\
\dot{\tilde{d}}_{4} & = & i(\Delta_{1}+\Delta_{2})\tilde{d}_{4}-i(g_{1}\tilde{d}_{2}+g_{2}\tilde{d}_{3})
.\label{t1}
\end{eqnarray}
We note the remaining equations that would arise by including the
possibility of two quanta per site form two independent sets, namely
\begin{eqnarray}
\dot{\tilde{d}}_{5} = -i\sqrt{2}g_{1}\tilde{d}_{6},\quad 
\dot{\tilde{d}}_{6} = i\Delta_{1}\tilde{d}_{6}-i\sqrt{2}g_{1}\tilde{d}_{5},
\end{eqnarray}
and 
 \begin{eqnarray}
\dot{\tilde{d}}_{7} = -i\sqrt{2}g_{2}\tilde{d}_{8},\quad
\dot{\tilde{d}}_{8} =  i\Delta_{2}\tilde{d}_{8}-i\sqrt{2}g_{2}\tilde{d}_{7},
\end{eqnarray}
 where states $5-8$ are $|\uparrow\downarrow10\rangle$, $|\downarrow\downarrow20\rangle$,
$|\downarrow\uparrow01\rangle$, $|\downarrow\downarrow02\rangle$
respectively. We have removed the fast oscillating terms by introducing
a rotating frame through the relation \begin{eqnarray}
\tilde{d}_{i}=e^{-i(\omega_{1}+\omega_{2})t}d_{i},\quad i=1,2,\ldots,8.\end{eqnarray}
 As before for the single excitation, we have allowed for the possibility
of imperfect couplings of the atoms to the cavity modes by introducing
detunings $\Delta_{1}$ and $\Delta_{2}$, and unequal coupling strengths
$g_{1}$ and $g_{2}$.

Note that the dynamics of the coupled amplitudes $d_{5}(t)$ and $d_{6}(t)$
as well as $d_{7}(t)$ and $d_{8}(t)$ involve, in fact, an exchange
of a single photon between the atoms and the cavity modes to which
they are coupled. They can display an entanglement during the evolution,
\textcolor{black}{but the dynamics will be similar to that discussed
in Sec. III for a single excitation}\textcolor{red}{.} Therefore,
we will not consider the evolution of these states and will focus
instead only on the evolution of the other two-photon states that
involve simultaneous evolution of two excitations.

Although the set of equations of motion (\ref{t1}) involves four
coupled equations, it is analytically solvable. We can put Eqs. (\ref{t1})
into a matrix form and solve them by the matrix inversion. Due to
the complexity of the general solution, we study separately the time
evolution of the probability amplitudes for two cases: (1) non-zero
detunings and equal coupling strengths, (2) zero detunings and unequal
coupling strengths.

Instead of the four equations for $\tilde{d}_{i}\ (i=1,2,3,4)$, we introduce the following combinations of the probability amplitudes
\begin{eqnarray}
d^{(1)}(t) & = & \tilde{d}_{1}(t)+\tilde{d}_{4}(t),\ d^{(2)}(t)=\tilde{d}_{1}(t)-\tilde{d}_{4}(t),\nonumber \\
d^{(3)}(t) & = & \tilde{d}_{2}(t)+\tilde{d}_{3}(t),\ d^{(4)}(t)=\tilde{d}_{2}(t)-\tilde{d}_{3}(t),
\end{eqnarray}
and find that in the first case, the time evolution of the amplitudes is given by 
\begin{eqnarray}
d^{(1)}(t) & = & d^{(1)}(0)\cos(\Omega gt)
 -\frac{2i}{\Omega}\left[d^{(3)}(0)+\delta d^{(2)}(0)\right]\sin(\Omega gt),\nonumber \\
d^{(2)}(t) & = & \frac{4}{\Omega^{2}}\left[d^{(2)}(0)-\delta d^{(3)}(0)\right]
 +\frac{4\delta}{\Omega^{2}}\left[d^{(3)}(0)+\delta d^{(2)}(0)\right]\cos(\Omega gt)\nonumber \\
 &  & -\frac{2i\delta}{\Omega}d^{(1)}(0)\sin(\Omega gt),\nonumber \\
d^{(3)}(t) & = & \frac{4\delta}{\Omega^{2}}\left[\delta d^{(3)}(0)-d^{(2)}(0)\right]
 +\frac{4}{\Omega^{2}}\left[d^{(3)}(0)+\delta d^{(2)}(0)\right]\cos(\Omega gt)\nonumber \\
 &  & -\frac{2i}{\Omega}d^{(1)}(0)\sin(\Omega gt),\nonumber \\
d^{(4)}(t) & = & d^{(4)}(0),\label{t5}\end{eqnarray}
 where the detunings are simplified to $\Delta_{1}=\Delta_{2}\equiv\Delta$,
so we have introduced a dimensionless Rabi frequency, the same for
both subsystems, $\Omega=2(1+\delta^{2})^{1/2}$ with~$\delta=\Delta/g$.

For the second case of exact resonance but unequal coupling strengths,
we find \begin{eqnarray}
d^{(1)}(t) & = & d^{(1)}(0)\cos(2gt)-id^{(3)}(0)\sin(2gt),\nonumber \\
d^{(2)}(t) & = & d^{(2)}(0)\cos(2ut)+id^{(4)}(0)\sin(2ut),\nonumber \\
d^{(3)}(t) & = & d^{(3)}(0)\cos(2gt)-id^{(1)}(0)\sin(2gt),\nonumber \\
d^{(4)}(t) & = & d^{(4)}(0)\cos(2ut)+id^{(2)}(0)\sin(2ut),\label{t6}\end{eqnarray}
 where, as before for one-photon states, $g=(g_{1}+g_{2})/2$ and
$u=(g_{1}-g_{2})/2$.

\subsection{Entanglement evolution for the double-excitation state}

\begin{figure}[hbp]
\begin{center}
\includegraphics[clip,width=4in]{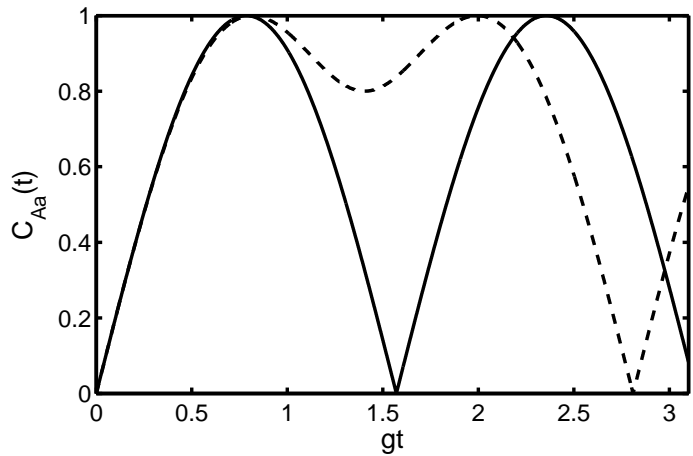}
\end{center}
\caption{Evolution of the concurrence $C_{Aa}(t)$ for the initial separable
state~$|\!\uparrow\uparrow\!00\rangle$ and two different detunings: $\Delta=0$
(solid line) and $\Delta=g$ (dashed line). 
\label{cap:localcouplingfig8}}
\end{figure}

We first investigate the evolution of the bipartite entanglement with
two quanta present in the system, but where there is always one excitation
in each of the subsystems $1$ and~$2$. The initial state of the
system is a linear superposition of the form 
\begin{eqnarray}
|\Psi\rangle =  d_{1}(t)|\!\uparrow\uparrow00\rangle+d_{2}(t)|\!\uparrow\downarrow01\rangle
 +d_{3}(t)|\!\downarrow\uparrow10\rangle+d_{4}(t)|\!\downarrow\downarrow11\rangle .
 \label{eq:twophotonc}
 \end{eqnarray}
This state cannot give any entanglement in the atom-atom subsytem,
regardless of the values of the coefficients involved. This is readily
seen by evaluating the reduced density operator for the atom systems.
We evaluate 
\begin{eqnarray}
\rho_{AB}={\rm Tr}_{cavity}\rho=\sum_{i,j=0,1}\langle i|\langle j|\rho|j\rangle|i\rangle,
\end{eqnarray}
where $|i\rangle$ and $|j\rangle$ refer to cavity modes $a$ and $b$ respectively, to get 
\begin{eqnarray}
\rho_{AB} = |d_{1}(t)|^{2}|\uparrow\uparrow\rangle+|d_{2}(t)|^{2}|\uparrow\downarrow\rangle
+|d_{3}(t)|^{2}|\uparrow\downarrow\rangle+|d_{4}(t)|^{2}|\downarrow\downarrow\rangle,
\end{eqnarray}
which is a mixture of separable states and hence cannot be entangled.
The atom-atom bipartite entanglement is always zero, $C_{AB}(t)=0$.
The same conclusion applies to the bipartite entanglements $C_{ab}(t),C_{Ab}(t)$
and $C_{Ba}(t)$. The only non-zero concurrence possible is for the
qubit pairs $C_{Aa}(t)$ and $C_{Bb}(t)$. In other words, an entanglement
can be created during the evolution between the atoms and the cavity
modes to which they are coupled. The amount of entanglement that can
be created in these pairs can be determined from the relations 
\begin{eqnarray}
C_{Aa}(t)=2|\tilde{d}_{1}(t)\tilde{d}_{3}^{\ast}(t)+\tilde{d}_{2}(t)\tilde{d}_{4}^{\ast}(t)|,\label{te1}
\end{eqnarray}
 and \begin{eqnarray}
C_{Bb}(t)=2|\tilde{d}_{1}(t)\tilde{d}_{2}^{\ast}(t)+\tilde{d}_{3}(t)\tilde{d}_{4}^{\ast}(t)|.\label{te2}
\end{eqnarray}
Figure \ref{cap:localcouplingfig8} shows the entanglement creation
and evolution between the atom $A$ and the cavity mode~$a$, with
an initial separable state $|\!\uparrow\uparrow00\rangle$. Similar
entanglement properties appear between the atom $B$ and the cavity
mode~$b$. We see that an entanglement between the atom and the cavity
mode is created during the transfer process of the excitation between
them. The maximum entanglement is observed at times when the excitation
is equally shared between the atom and the cavity mode, and vanishes
when the excitation is completely located at either the atom or the
cavity mode. In terms of the Rabi oscillations, the system becomes
disentangled at every half of the Rabi cycle of the oscillation, i.e.
at $gt=n\pi/\Omega\ (n=0,1,2,\ldots)$.

Similar to the case of single excitation states of Section~\ref{sec3}, the
system becomes entangled for a longer time when the frequencies of
the atom and the cavity field are detuned from each other. In other
words, an imperfect matching between the atom and the cavity field
leads to a more stable entanglement than in the case of the perfect
matching. In terms of the Rabi oscillations, the system becomes disentangled
only at every Rabi cycle of the oscillation, i.e. at $gt=2n\pi/\Omega$.
This is easy to understand if one considers entanglement as resulting
from a superposition corresponding to a delocalization of energy.
When the frequencies of the atom and the cavity mode are different,
the initial amount of energy localized in the atom is not completely
transferred to the cavity mode. A part of the energy, not absorbed
by the cavity mode remains delocalized, which results in a nonzero
entanglement.

\subsection{Inclusion of the auxiliary state $|0\rangle$ }

The above analysis has showed that no pairwise entanglement is possible
between any parts of the two subsystems $1$ and $2$ when we have
with certainty a single excitation in each of $1$ and $2$. We can
introduce entanglement by including in the wave function an auxiliary
state, the ground state $|0\rangle$ for which there is no excitation~\cite{yonac}: 
\begin{eqnarray}
|0\rangle=|\!\downarrow\downarrow00\rangle.\label{es0}
\end{eqnarray}
The evolution of the entire wave function can be written explicitly as 
\begin{eqnarray}
|\Psi\rangle & = & d_{1}(t)|\!\uparrow\uparrow00\rangle+d_{2}(t)|\!\uparrow\downarrow01\rangle
+d_{3}(t)|\!\downarrow\uparrow10\rangle+d_{4}(t)|\!\downarrow\downarrow11\rangle\nonumber \\
 &  & +d_{0}(t)|\!\downarrow\downarrow00\rangle.\label{eq:twophotonc0}
 \end{eqnarray}

It is easy to show that the state (\ref{es0}) is not the eigenstate
of the Hamiltonian (\ref{e1}). Therefore, the probability amplitude
$d_{0}$ does not evolve in time, $d_{0}(t)=d_{0}(0)$. As before
for the single-excitation states, we will examine the entanglement
between different pairs of the subsystems and its transient behaviour
especially under imperfect matching conditions. For this general case,
the pairwise concurrences between the subsystems are found to be 
\begin{eqnarray}
C_{AB}(t) & = & \max\bigl\{0,2(|d_{1}(t)||d_{0}(t)|-|d_{2}(t)||d_{3}(t)|)\bigr\},\nonumber \\
C_{ab}(t) & = & \max\bigl\{0,2(|d_{4}(t)||d_{0}(t)|-|d_{2}(t)||d_{3}(t)|)\bigr\},\nonumber \\
C_{Ab}(t) & = & \max\bigl\{0,2(|d_{2}(t)||d_{0}(t)|-|d_{1}(t)||d_{4}(t)|)\bigr\},\nonumber \\
C_{aB}(t) & = & \max\bigl\{0,2(|d_{3}(t)||d_{0}(t)|-|d_{1}(t)||d_{4}(t)|)\bigr\}. \label{eq:conc2}
\end{eqnarray}
 Evidently, we see that entanglement between the pairs of subsystems is possible \emph{only} if the ground state is included.

We note interesting properties of the concurrence and differences
between the single and double-excitation cases. Firstly, the concurrences
appear as differences of products of the absolute values of the probability
amplitudes. This gives a possibility for discontinuities in the behavior
of entanglement that was absent in the case of the one-excitation,
where the concurrences, given by~(\ref{eq:conds}), are continuous
functions. The abrupt disappearance of entanglement that persists
for a period of time is referred to as {}``sudden death of entanglement''
(ESD) and was reported by Yonac et al.~\cite{yonac} for this case.
Secondly, there is evident a competition in the creation of entanglement
$C_{AB}$ between two pairs of states $(|\Phi_{1}\rangle,|\Phi_{4}\rangle)$
and $(|\Phi_{2}\rangle,|\Phi_{3}\rangle)$. Entanglement creation
in concurrence pairs involving the states $|\Phi_{1}\rangle$ and
$|\Phi_{4}\rangle$ is diminished by the presence of population in
the states $|\Phi_{2}\rangle$ and $|\Phi_{3}\rangle$, and vice versa
creation of entanglement involving the states $|\Phi_{2}\rangle$
and $|\Phi_{3}\rangle$ is diminished by the presence of population
in the states $|\Phi_{1}\rangle$ and $|\Phi_{4}\rangle$. In terms
of the population transfer, entanglement creation by a simultaneous
exchange of two photons is diminished by one-photon exchange processes,
and vice versa. The competition between these one and two-photon processes
has been recognized as the source of the phenomenon of ESD, sudden
death of entanglement.

\subsection{Inequality for sum of square of the nonlocal pairwise concurrences:}

The reduced density matrix for the atom-atom system, in terms of basis
states $|\!\uparrow\uparrow\rangle$, $|\!\uparrow\downarrow\rangle$,
$|\!\downarrow\uparrow\rangle$ and $|\!\downarrow\downarrow\rangle$, is 
\begin{eqnarray}
\rho_{AB} & = & \left(\begin{array}{cccc}
|d_{1}|^{2} & 0 & 0 & d_{1}d_{0}^{*}\\
0 & |d_{2}|^{2} & 0 & 0\\
0 & 0 & |d_{3}|^{2} & 0\\
d_{1}^{*}d_{0} & 0 & 0 & |d_{4}|^{2}+|d_{0}|^{2}\end{array}\right),\label{eq:densitymat}
\end{eqnarray}
for which the concurrence is $C_{AB}=2(|d_{1}||d_{0}|-|d_{2}||d_{3}|)$.
We drop the explicit time dependence for simplicity of notation.
\begin{figure}[h]
\begin{center}
\includegraphics[clip,width=4in]{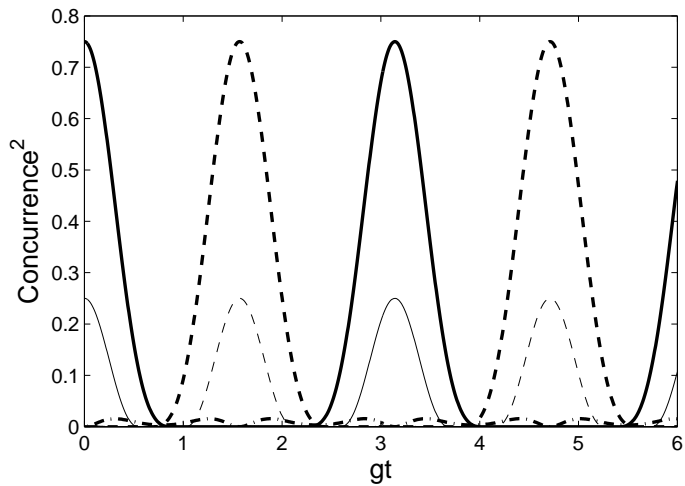}
\end{center}
\caption{Evolution of entanglement for the Bell state $\Phi$ showing the lack of conservation
of the sum of the square of the nonlocal pairwise $1-2$ concurrences, 
$SSPC=|C_{AB}|^{2}+|C_{Ab}|^{2}+|C_{aB}|^{2}+|C_{ab}|^{2}$.
\textcolor{black}{Here $\Delta=0$ and $g_{1}=g_{2}=g$}. The squares of concurrences are plotted in bold for $\alpha=\pi/6$, and as thin lines, for $\alpha=\pi/12$. We have $C_{AB}^{2}$ (solid line), $C_{ab}^{2}$
(dashed line), $C_{Ab}^{2}=C_{aB}^{2}$ (dash-dotted line). As the initial entanglement
weakens, to $\alpha=\pi/12$ for example, regions of total loss of all non-local pairwise entanglement are evident
(SSPC=0). Total pairwise $1-2$ entanglement (SSPC) is restored when the
entanglement has been completely transferred from atoms to cavity modes. 
\label{cap:pairwiseloss9ab}}
\end{figure}

In contrast with the single-excitation case of Section~\ref{sec3}, there
is the difference that the reduced state can be entirely separable,
for periods of time (sudden death), even when the off-diagonal coherence
term $d_{1}d_{0}^{*}$ which is the source of the entanglement is
non-zero.\textcolor{black} This feature results in a reduction of
the sum total of the squares of the nonlocal pairwise concurrences (SSPC) and is also
responsible for the entanglement sudden death (ESD) phenomenum. The entanglement
for the remaining reduced systems, the two cavities, and atom and
cavity at different location, is given similarly, by the concurrences
as in~(\ref{eq:conc2}).

In fact, for the Bell state $\Phi$ of the form (\ref{eq:bell2}),
we can identify regions of evolution for which each of $C_{AB}$,
$C_{\alpha b}$, $C_{aB}$, $C_{ab}$ is zero, meaning that all nonlocal pairwise
entanglement involving systems $1$ and $2$ is lost.
This effect has been discussed by Chan et al~\cite{smznew}, and is evident in 
Fig.~\ref{cap:pairwiseloss9ab}, for $\alpha=\pi/12$.
In this case, the entanglement involves the four states, that of the
atoms and cavity modes. We readily see from~(\ref{eq:conc2})
that where we have $|d_{1}|=|d_{2}|=|d_{3}|=|d_{4}|=\frac{1}{2}$,
for any nonzero value of $d_{0}=\sin\alpha$, each pairwise concurrence
between systems $1$ and $2$ will be zero. The entangled state at
$t=n\pi/4\Omega$, where $n=1,3,5...$ , that can give this total
depletion of all pairwise entanglement for $\tan\alpha<1/2$, is written
\begin{eqnarray}
\frac{1}{2}\cos\alpha(|\!\uparrow\uparrow00\rangle+|\!\uparrow\downarrow01\rangle+|\!\downarrow\uparrow10\rangle+|\!\downarrow\downarrow11\rangle)
+\sin\alpha|\!\downarrow\downarrow00\rangle.\label{eq:statezeroconc}
\end{eqnarray}

\subsection{Controlling ESD using detunings}

\begin{figure}[h]
\begin{center}
\includegraphics[clip,width=4in]{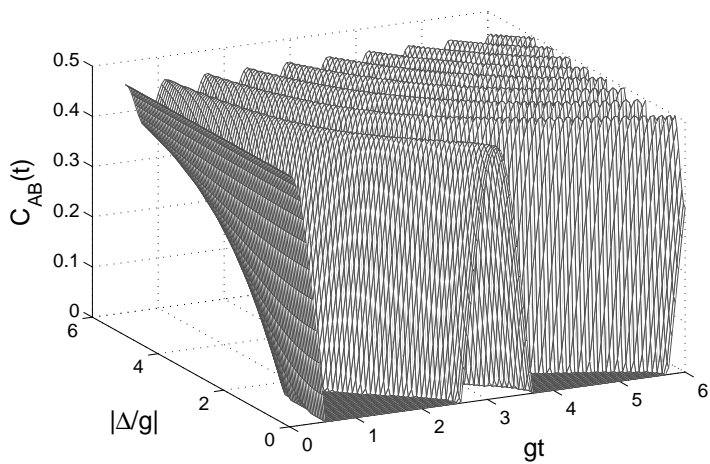} 
\end{center}
\caption{Plot of the evolution of $C_{AB}$ for the intial Bell state
$\Phi$, given by~(\ref{eq:bell2}). Here $g=g_{1}=g_{2}$ and $\alpha=\pi/12$.
The effect of detuning $\Delta\neq0$ is to remove the ESD, the sudden
death of entanglement. 
\label{cap:detuningesdfig10}}
\end{figure}

Sudden death of entanglement has been reported by Yonac et al.~\cite{yonac},
for the case where the initial state is the Bell state $\Phi$ 
\begin{equation}
|\Phi\rangle=\cos\alpha|\uparrow\uparrow00\rangle
+\sin\alpha|\downarrow\downarrow00\rangle ,\label{eq:bell222}
\end{equation}
and where the system has zero detunings and equal coupling constants.
They showed that for $\alpha\neq\pi/4$, entanglement is lost in a
discontinuous fashion, as displayed in Fig. \ref{cap:detuningesdfig10},
\textcolor{black}{for $\alpha=\pi/12$. As} $\tan\alpha$ deviates
from $1$, the length of time for which entanglement is lost is increased.

We observe from the full solutions that an imperfect matching of the
cavities to the atomic transition can prevent the onset of ESD, to
provide a stabilization of entanglement. This effect, which is similar
to that of Figure \ref{cap:detuningcon5}, is shown in Fig.~\ref{cap:detuningesdfig10},
and may be understood by realizing that the nonzero detuning $\Delta$
inhibits the full entanglement transfer from atom to cavity modes.

For some initial states however, we point out that the detuning will
act to induce ESD, that is, to destabilise the entanglement, though
stabilization is reached for sufficiently large $\Delta$. To further
investigate the role of the imperfect matching on the evolution of
entanglement, we consider two different initial states, one that displays
ESD with zero detuning $\Delta$, and another that displays stable
entanglement with $\Delta=0$. We consider a linear superposition
of the zero-excitation state with a two-quanta symmetric state 
\begin{equation}
\frac{1}{\sqrt{2}}\left(\frac{1}{\sqrt{2}}(|\uparrow\uparrow00\rangle
+|\downarrow\downarrow11\rangle)+|\downarrow\downarrow00\rangle\right),\label{eq:sym_state_1}
\end{equation}
 and a linear superposition of the zero excitation state with a two-quanta
antisymmetric state 
\begin{equation}
\frac{1}{\sqrt{2}}\left(\frac{1}{\sqrt{2}}(|\uparrow\uparrow00\rangle
-|\downarrow\downarrow11\rangle)+|\downarrow\downarrow00\rangle\right) 
.\label{eq:ant_sym_state_1}
\end{equation}

\begin{figure}[h]
\begin{center}
\includegraphics[clip,width=4in]{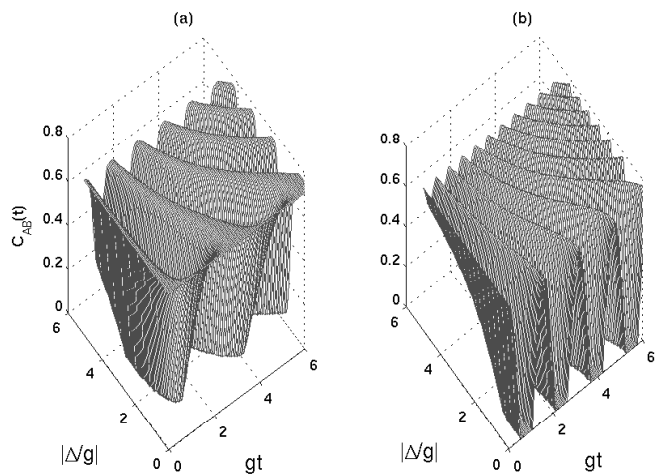} 
\end{center}
\caption{Concurrence between the atoms as a function of time $t$ and detuning
$\Delta$ for two different initial states: (a) the asymmetric state~(\ref{eq:ant_sym_state_1}) 
and (b) symmetric state~(\ref{eq:sym_state_1}).
\label{cap:detuningsstatesfig11ab}}
\end{figure}

Figure \ref{cap:detuningsstatesfig11ab} shows the evolution of concurrence
$C_{AB}(t)$ as a function of the detuning~$\Delta=\Delta_{1}=\Delta_{2}$
of the cavity modes from the atomic resonances. We see from the figure
that the detuning has an opposite effect on the two initial states:
the stable entanglement at $\Delta=0$ for the antisymmetric state
is forced into an entanglement sudden death (ESD) with the onset of
a non-zero $\Delta$, whereas the unstable entanglement evolution
(ESD) displayed by the symmetric state at $\Delta=0$ is stabilized
by detunings $\Delta\neq0$. We note in both cases however, that for
very large $\Delta$, the entanglement is stabilized at or near the
intial value.

\subsection{Effect of asymmetric coupling constants}

\begin{figure}[h]
\begin{center}
\includegraphics[width=4in]{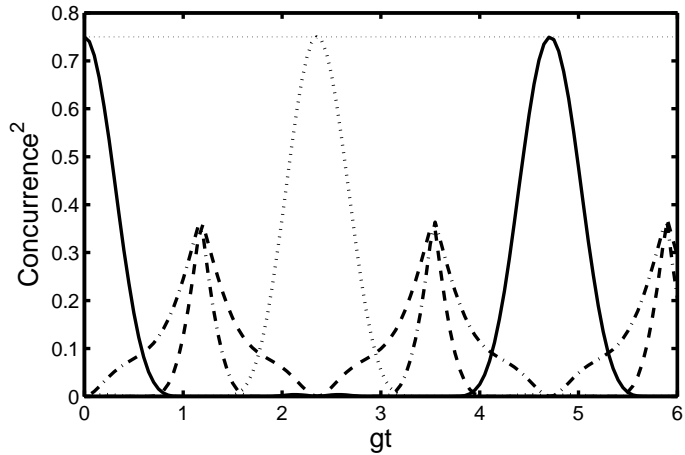}
\end{center}
\caption{Evolution of the nonlocal pairwise concurrences for the Bell state $\Phi$,
where couplings are asymmetric\textcolor{black}{: $g_{2}=2g_{1}$,
$\Delta=0$. Here $\alpha=\pi/6$. The en}tanglement sudden death
(ESD) is evident, as is the lack of conservation of the sum of the
square of the pairwise $1-2$ concurrences, 
$SSPC=|C_{AB}|^{2}+|C_{Ab}|^{2}+|C_{aB}|^{2}+|C_{ab}|^{2}$.
Individual concurrences are: $C_{AB}^{2}$ (solid line), $C_{ab}^{2}$
(dashed line), $C_{aB}^{2}$ (dash-dotted line), $C_{Ab}^{2}$ (dotted
line). Also plotted is \textcolor{black}{ $C_{12}^{2}$ , the horizontal
dotted line at $0.75$}. 
\label{cap:asymgssc}}
\end{figure}

\begin{figure}[hbp]
\begin{center}
\includegraphics[width=4in]{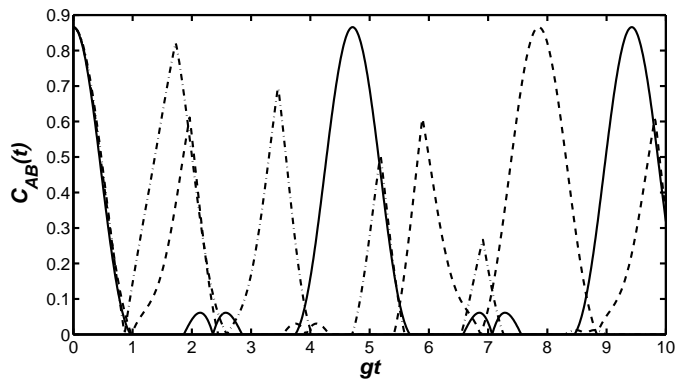}
\end{center}
\caption{Evolution of atom-atom concurrence for the Bell state $\Phi$,
with $\alpha=\pi/6$, where couplings \textcolor{black}{are asymmetric:
$\Delta=0$ and $g_{1}=g_{2}$ (solid line); $g_{2}=2g_{1}$ (dashed
line); $g_{2}=10g_{1}$ (dashed-dotted line). The ESD shifts toward
a later time as the ratio of coupling constants increases. }\label{cap:asymgesd}}
\end{figure}

The loss of pairwise entanglement for the spin correlated Bell state
$\Phi$ is a feature maintained even in the case of asymmetric couplings:
$g_{1}\neq g_{2}$. The depletion of entanglement (ESD) for the atom-atom
pair has been reported, for asymmetric couplings, by Sainz and Bjork\textcolor{black}{~\cite{sainz}.
In Fig. \ref{cap:asymgssc}, we plot the non-local pairwise concurrences from
which one can calculate the sum of their squares, $SSPC$, as in  ~(\ref{eq:concurrenceunequal}),
to reveal a loss of total pairwise entanglement where couplings are
asymmetric. As with the symmetric case, the effect is more pronounced
with a decreasing initial entanglement, $C_{12}$. We note from Fig.
\ref{cap:asymgesd} that there is a shift of the period of entanglement
sudden death, as the ratio of the $g$'s increases.}

\begin{figure}[hbp]
\begin{center}
\includegraphics[clip,width=4in]{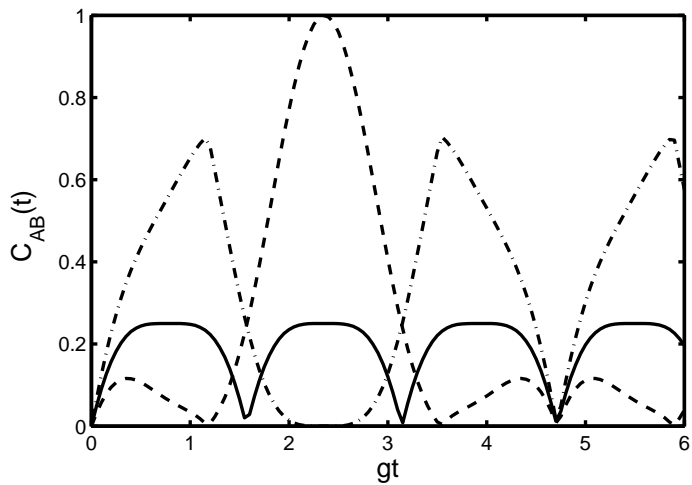} 
\end{center}
\caption{Evolution of the concurrence $C_{AB}(t)$ for the initial state
$|\Lambda\rangle$ with  $g_{1}=g_{2}$ (solid line), $g_{2}=2g_{1}$
(dashed line) and $g_{2}=0.5g_{1}$ (dashed-dotted line). 
\label{cap:asymgesdfig13}}
\end{figure}

In Fig. \ref{cap:asymgesdfig13}, we plot the atom-atom concurrence
$C_{AB}(t)$, for different coupling strengths, for the different
initial state
\begin{equation}
|\Lambda\rangle=\frac{1}{\sqrt{2}}(|\!\downarrow\uparrow10\rangle
+|\!\downarrow\downarrow00\rangle) ,\label{eq:lambda state}
\end{equation}
which has no entanglement between the two atoms.
In contrast to the state $\Phi$, there is no abrupt depletion (ESD)
of entanglement. We note that a suitable choice of coupling ratio
can lead to a periodic maximal entanglement bewteen the atoms. 

In summary of this section, we point out that small differences in
the frequencies can lead to the disappearance of the zeros (ESD) in
the entanglement evolution, whereas small differences in the coupling
strengths do not generally prevent the vanishing of entanglement,
but rather tend to create ESD. In other words, entanglement sudden
death is much more sensitive to an imperfection in the frequencies
than to the coupling strengths.

\section{Conclusion}

We have studied the time evolution of entanglement between two remote
and in general non-identical Jaynes-Cummings systems. A particular
attention has been paid to the effect of an imperfect matching of
the atoms to the cavity fields and to the understanding of the nature
of entanglement evolution and steered transfer. We have quantified
entanglement through the pairwise concurrence between different parts
of the two systems and have found that entanglement transfer can be
triggered and controlled by a suitable choice of the detunings between
the atomic and the cavity field frequencies and the coupling strengths
of the atoms to the cavity fields. We consider two cases: the first
where there is only one excitation present in the entire system; the
second where there is a superposition of two excitation levels, zero
excitation and two excitations, one in each of the two remote systems.
These two cases exhibit completely different properties of the entanglement
evolution and steering of the entanglement transfer. When the atomic
and field frequencies are detuned from each other and there is only
one excitation present, an initial entanglement cannot be completely
transferred between the pairwise subsystems. On the other hand, for
zero detunings one can achieve complete transfer of the entanglement
to a desired pair of the subsystems and this process can be steered
by a suitable choice of the ratio between the coupling strengths.
In the second case where two excitations are present in the system,
we have showed that no entanglement is possible between any parts
of the two subsystems when the number of photons in each subsystem
is conserved. An additional auxiliary state of zero excitation has
to be included to create entanglement between the subsystems. We also
discuss a conservation rule governing the pairwise entanglement between
the systems, that the sum of the squares of the pairwise concurrences
is conserved during the evolution of an initial entanglement, for
the one-excitation case. For the superposition of the zero and two
excitation case, the sum is not conserved, and can be zero for intervals
of time, implying an entanglement sudden death in all pairwise concurrences.
Finally, we have addressed the issue of sudden death of entanglement
in relation to non-identical cavities, and have found that a mismatch
in the frequencies can lead to the disappearance of sudden death in
the entanglement evolution, whereas differences in the coupling strengths
can prevent and even create the vanishing of entanglement.

\section*{Acknowledgement}

This research was supported by the Australian Research Council. One
of us, MDR, acknowledges support from the ARC Centre of Excellence
Program.


\section*{References}

\begin{thebibliography}{10}
\bibitem{expent} Turchette Q A, Wood C S, King B E, Myatt C J, 
 Leibfried D, Itano W M, Monroe C and D. J. Wineland  D J 1998 Phys. Rev.
Lett. \textbf{81} 3631 

\bibitem{atomicens} Julsgaard B, Kozhekin A and Polzik E S 2001 Nature
\textbf{413} 400

\bibitem{aspecttype} Aspect A, Grangier P and Roger G 1981 Phys. Rev.
Lett. \textbf{47} 460 

\bibitem{diosi} Diosi L 2003 Lect. Notes Phys. \textbf{622} 157

\bibitem{dodd} Dodd P J and Halliwell J J 2004 Phys. Rev. A {\bf 69} 052105

\bibitem{yueberly}  Yu T and  Eberly J H 2004 Phys. Rev. Lett. \textbf{93} 140404

\bibitem{ebscience} Eberly J H and Yu T 2007 Science \textbf{316} 555 

\bibitem{alexp} Almeida M P et al. 2007 Science \textbf{316} 579

\bibitem{salles} Salles A et al, 2008 Phys. Rev. A\textbf{78} 022322

\bibitem{zt}  Ficek Z and Tana\'s R 2006 Phys. Rev. A\textbf{ 74} 024304

\bibitem{santos} Santos M F, Milman P, Davidovich L and Zagury N 2006
Phys. Rev. A \textbf{73} 040305

\bibitem{other esdanna} Jamroz A 2006 J. Phys. A \textbf{39} 7727

\bibitem{recent esd terra}Terra Cunha M O 2007 New J. Phys. \textbf{9} 237

\bibitem{sainz} Sainz I and Bjork G 2007 Phys. Rev. A \textbf{76} 042313

\bibitem{caval} Cavalcanti D et al. 2006 Phys. Rev. A \textbf{74} 042328

\bibitem{esd} Al-Qasimi A and James D F V 2008 Phys. Rev. A\textbf{77} 012117

\bibitem{yebx} Yu T and Eberly J H 2007 Quant.  Inf. and Comp. \textbf{7} 459

\bibitem{esd2008} Tahira R,  Ikram M, Azim T and  Zubairy M S 2008 J. Phys. B: At. Mol. Opt. 
Phys. \textbf{41} 205501 and references therein.

\bibitem{qed work exp} Ye J, Vernooy D V and Kimble H J 1999 Phys. Rev. Lett. \textbf{83} 4987

\bibitem{har-ryd} Hagley E et al. 1997 Phys. Rev. Lett. \textbf{79} 1

\bibitem{jc} Jaynes E T and Cummings F W 1963 Proc. \emph{IEEE} \textbf{51} 89

\bibitem{jcexp} Brune M et al. 1996 Phys. Rev. Lett. {\bf 76} 1800

\bibitem{yonac} Yonac M, Yu T and Eberly J H 2006  J. Phys. B, At Mol Opt Phys. \textbf{39} S621

\bibitem{yonac2} Yonac M, Yu T and Eberly J H 2007 J. Phys. B, At Mol Opt Phys. \textbf{40} S45

\bibitem{yelattice} Yonac M and Eberly J H 2008  Opt. Lett. \textbf{33} 270

\bibitem{wit2}  Wootters K 1998 Phys. Rev. Lett. \textbf{80} 2248

\bibitem{smznew} Chan S, Reid M D and Ficek Z, to be published.

\bibitem{lopez} Lopez C E et al. 2008 Phys. Rev. Lett. \textbf{101} 080503
\end{thebibliography}
\end{document}